\documentstyle[12pt]{article}
\begin{document}
\title{ Classicality via hydrodynamics in quantum field theory} 
\author {C. Anastopoulos \\ Departament de Fisica Fonamental \\
Universitat de Barcelona \\ Av. Diagonal 647, 08028 Barcelona\\
 Spain \\
 E-mail: charis@ffn.ub.es \\}
\date { May 1998}
\maketitle
\begin{abstract}
Motivated by ideas from the consistent histories approach to quantum mechanics,
 we examine a simple model of hydrodynamic coarse graining for a scalar field.
 It consists in averaging the field over spatial regions of size $L$ and constructing the evolution equations for the coarse grained quantities, identifying dissipation and noise. 
\end{abstract}

 \renewcommand {\theequation}{\thesection.\arabic{equation}}
\let \ssection = \section
\renewcommand{\section}{\setcounter{equation}{0} \ssection}
\pagebreak

\section{Introduction}
\subsection{Motivation}
In modern cosmology there seems to exist a gap between our  description of early universe using essentially quantum field theory (hence potentially involving highly non-classical behaviour of matter) and the later time (up to current era) description of matter as essentially classical by a few phenomenological parameters (as corresponding for instance to a classical fluid).
\par
This gap is not conceptual: one can intuitively visualise that as the scale factor becomes larger somehow the quantum behaviour of the fields gets smoothed out and matter falling to lower energies can essentially be described by a collective set of variables, (usually hydrodynamic). The problem arises when one tries to give a more precise description of the underlying mechanisms of this process.
\par
It can be safely  said that the quantum to classical transition for quantum field theories has not been understood very well yet. This is in spite of the large amount of activity in recent years and the application of relatively recent concepts and ideas, successful in the realm of non relativistic quantum mechanics (the environment superselection rules approach \cite{Zur}, consistent/decoherent histories \cite{Gri,Omn,GeHa}). 
\par
The issue is more strongly highlighted in the context of inflationary models; a necessary ingredient in them is the amplification of quantum fluctuations to classical so that the inhomogeneity can be properly explained. But so far there is no consensus on how this classicalisation is taking place and more importantly why.
\par
The problem of course stems from the extremely complicated nature of quantum field theories. Being a systems with an infinite number of degrees of freedom there is a large (both technical and conceptual ) difficulty into implementing  schemes succesful in ordinary quantum mechanics. 
\par
What is usually considered in such discussions is the phase space classical limit of a quantum field, that is the emergence of a classical field limit from the full quantum field theory. In such a limit the expectation value of the field evolves according to the Hamilton equations plus the action of some stochastic terms. Since in such discussions usually free fields are considered, the discussion essentially comes into the consideration of each mode separately.
\par
The question of course exists whether the results coming from mode by mode splitting are robust enough when one takes into account field interaction. But even, letting this issue unanswered, there are is a number of problems. It is sometimes stated that modes become classical solely by the squeezing induced  throught the time variation of the scale factor. There are strong arguments against this interpretation: from a phase space perspective sueezing entails a highly non - classical behaviour (see \cite{Ann} for a more thorough discussion)
\par
Inspired by the success of the environment induced decoherence program the coupling of the field to some kind of environment has been proposed as a mechanism for classicality. There are some problems with this approach two: is there a natural splitting between system and environment in early universe? decoherence in non-relativistic quantum mechanics appears only when environment is much larger than the system; is this the case in field theory or spaecial initial conditions are to be assumed? Another idea professed in this context is that the role of the environment could be played by the higher order (inaccesible to the local observer ) n - point funtion sector of the field \cite{CaHu,Ana}. 
\par
This is an interesting possibility, which forms one of the two main motivating ideas of this paper. The other (and main one) is the Gell - Mann and Hartle 
decoherent histories program \cite{GeHa}. In this formalism, the classical limit is obtained by the study of coarse grained temporal propositions (histories) about the physical system. Classicality is equivalent to a consistency condition between pairs of histories, quantified by the vanishing of the off - diagonal terms of the decoherence functional.
\par
Concerning our aims, the great strength and versatility of this approach lies in the fact that it provides a generic algorithm to determine whether a set of histories effectively becomes classical or not, irrespective of the physical quantities to which they refer. Hence, in principle, it is sufficient to specify the suitable coarse grained operators, to be able to determine the existence of a corresponding quasiclassical domain. The discussion of classicality is therefore not restricted to the phase space properties of the system. 
\par
As we argued, in the case of field theory phase space clasicality is a rather difficult issue to study. But in order to obtain an effective, cosmologically relevant description it would be sufficient to concentrate on much more coarse - grained observables: essentially hydrodynamic or thermodynamic ones. Indeed, the emergence of hydrodynamic behaviour in many body systems (starting from first principles and no assumptions as local equilibrium) has been persistently advocated by Gell-Mann and Hartle. The issue is very important (not only in the cosmological setting), but involves such a degree of technical complexity, that only very simple systems have been examined so far \cite{HaBr}.
\par
In any case , we believe that concentration on hydrodynamic properties of quantum fields is an important and not sufficiently addressed avenue towards understanding the emergence of classicality in quantum field theory. A great potential advantage of this approach, is that {\it the observables involved are much more coarse-grained } than the phase space ones and hence classicality might be emergent even if the field does not have a good classical field limit.

{\bf This paper }
We are not in a position to give a full treatment of the emergence of hydrodynamic behaviour in field theory, within or without the consistent histories framework. Our aim in this paper is much more modest: in the simplest possible field theory (scalar field in Minkowski spacetime) we shall try to see how a simple hydrodynamic type of coarse graining can be implemented, focusing mainly on the general quantitative features of the evolution of coarse grained quantities. We should stress here, that even in such a simple system the behaviour is anything but trivial.
\par
But before explaining our model in more detail, let us first give some general considerations.
\par
Any hydrodynamic description of a theory, whose full dynamics are   known 
 involves three spatial scales, a clean separation of which is usually necessary to render the 
description be succesful . The first one, $l_{mic}$ is the time-scale within which 
microscopic processes are taking place. This is essentially determined both by the dynamics and the initial state of the system. The second scale, $l_{av}$ is the scale, within which microscopic processes are averaged out. A volume of the size of $l_{av}^3$ can be thought as being sufficiently and independently described by a spatial average of all relevant physical quanties. This scale is determined primarily by the previous history of the system, and at least in the field theory case its existence 
is thought to necessitate a particular class of initial states. It is clear that we should have
$l_{mic} << l_{av}$.
 Finally, the third scale $l_{obs}$ corresponds to the level of observation. It is fixed by the 
external constraints to the system . If one wants to take the continuum approximation, one necessarily has to demand that $l_{av} << l_{obs}$, but even if this is not true an effective lattice type of description can be implementd for the coarse grained dynamics.
\par
These consideations  fix more or less the strategy we shall adopt in this paper. We are going to consider a coarse-graining operation for the field consisting in spatial averaging within a spatial region of volume  $L^3$. That is, we consider a foliation of spacetime and to each Cauchy hypersurface we construct a cubic lattice of side $L$ (which is to be identified with $l_{av}$)
. The field averaged values are to be thought as our coarse grained variables and an effective equation giving the evolution of their expectation values is going to be  constructed through the use of the Zwanzig projector technique \cite{Ana, ReMi, Zeh}. This is in spirit similar to a technique frequently employed in astrophysical discussions.
\par
A number of remarks are necessary at this point. First of all our choice for coarse- graining is not Lorenz invariant. Clearly, a different choice of foliation would lead to different value for $L$ and hence to different expressions for the evolution equations. In particular the condition 
$l_{mic} << l_{av}$ might not hold in a new Lorentz frame. It is in this sense that both our choice of foliation and the value of $L$ is not to be thought asarbitrary, but rather as phenomenological, and contingent upon a particular (cosmological) history. This is to be understood in the following way: Suppose that at some stage of the cosmological evolution, we observe that there is a preferred class of hypersurfaces (possibly the surfaces of approximate isotropy), such that: \\
1. Histories of field variables smeared in regions of volume $L^3$ decohere \\
2. $l_{mic}$ as can be read from the dynamics and the initial state (essentially would correspond to the maximum of characteristic length scales in the decoherence functional) 
 is much smaller than 
$L$. \\
3. Our scale of observation (which is to be thought in the order of magnitude of the scale factor) is much larger than $L$. \\
Then the evolution equations for this coarse-grained field variables, can essentially be constructed along the lines we shall develop in this paper. We cannot help overemphasizing, that the value for $L$ is not arbitrary. Rather we should consider it to correspond to the minimum value for which histories of averaged field variables decohere, and should be computed from first principles did we have a satisfactory (and computationally tractable) history version of field theory (or any other classicalisation scheme) as well as a good specification of the initial condition
\par
We believe, that an analogy with low energy physics wold help to clarify the above discussion.
Consider a box ( whose side we take to correspond to $l_{obs}$), in which at one paricular
 moment of time,  we insert a large number of nuclei  and electrons. The length scale of interaction for this highly active gas, should be decided by weighting  a mean of the de-Broglie wavelengths of the particles ( which is determined by the initial state) with some possible length-scale determined by electromagnetic screening (hence the Hamiltonian is also contributing).  Given sufficient time (and assuming low initial energy), our experience says that clusters will form (atoms)  the evolution of which will be substantially less ``violent''. Now it may happen as we focus our attention to a particular region of the box (with volume $L^3$), the average value of physical quantities, will change smoothly over time and will not exhibit irregularities of quantum mechanical nature. Hence to these averaged variables we shall be able to assign probabilities. Taking the smallest value for $L$ for which this is possible, we shall have the possibility of 
giving a description of the system in terms only of such coarse-grained observables. If in addition, $L$ turns out to be much smaller than the size of the box, then a continuity assumption can be invoked and we would finally arrive at a hydrodynamic description of the system. For this regime, in principle, we should be able to construct a Navier-Stokes type of equation.
\par
As we said we shall restrict ourselves to a free field case. Independently of the above considerations, this is of interest on its own right: in recent work on stochastic inflation \cite{Mat} coarse graining of the inflaton field has been considered corresponding essentially to  smearing the field in horizon scale spatial volume. In such coarse graining, it was assumed that only the low energy modes significantly contribute. In our model we shall be able to verify the truth of such an assertion, as well identify a source of noise due to the long range correlations of the initial state which is overlooked in such discussions.
\par
 We should add  that actual decoherence of field variables  seems to necessitate interaction terms, but as soon as 
this is established, the quadratic part of the Hamiltonian is the one giving the dominant contribution to the hydrodynamic equations (at least in the perturbation theory regime). The reason for this is that the coarse graining operation does not commute with the Hamiltonian, and even if interaction terms are assumed, their contribution is going to be of the next order in perturbation theory. In any case, there is no conceptual or technical difficulty in considering interaction terms ( this can be done by the conjuction of our coarse graining operator with the one of \cite{Ana}).

\section{Implementing the coarse - graining}

\subsection{ The formalism}
\paragraph{The Zwanzig method} 
As we mentioned in the introduction , we are going to employ the Zwanzig projection formalism for the implementation of the coarse - graining \cite{Ana, ReMi, Zeh}. We remind the reader that this consists in introducing an indempotent
``coarse graining operator'' in the space of states
\begin{equation}
\rho \rightarrow \rho_{rel} =  {\bf P} \rho \hspace{2cm} {\bf P}^2 =
{\bf P}
\end{equation}
which projects the state in the level of description.
The irrelevant part of the state is then given by 
\begin{equation}
\rho_{irr} = ({\bf 1} - {\bf P}) \rho
\end{equation}
By duality through the trace functional a corresponding  operator can be 
introduced in the space of observables and we can obtain the evolution equation for the expectation values of relevant observbles i.e. the ones for which ${\bf P} A = A$ ( ${\bf P}$  is assumed self - adjoint)

\begin{eqnarray}
i \frac{\partial}{\partial t} \langle A \rangle(t) -  \langle
{\bf PL} A \rangle(t) \hspace{3cm} \nonumber \\
+ i \int_0^t d \tau \langle {\bf PL}({\bf
1}- {\bf P}) e^{i  {\bf L} \tau} ({\bf
1}- {\bf P}) {\bf LP} A \rangle(t-\tau) = F_A(t)
\end{eqnarray}
where ${\bf L} A = [H,A]$ in terms of the Hamiltonian $H$
and $e^{-i {\bf L}t}$ is the corresponding evolution operator.
\par
$F_A(t)$ is a  stochastic in nature term, 
since it
depends on the irrelevant components of the initial state that are
inaccesible from our level of description. It reads
\begin{equation}
 F_A(t) = - Tr \left( \rho(0){\bf F}(t) A \right)
\end{equation}
where 
\begin{equation}
{\bf F}(t) = ({\bf 1} 
- {\bf P}) e^{i  {\bf L} t} ({\bf 1} 
- {\bf P}){\bf L}
\end{equation}
This kernels is sufficient to determine the correlator of the noise in term of the n - point functions of the initial state since
\begin{equation}
\langle F_A(t) F_B(t') \rangle = Tr \left( \rho(0) ({\bf F}(t) A) ({\bf F}(t) B) \right)
\end{equation} 
\paragraph{The free field} We have found more convenient to use the Fock representation for the free field. Our basic objects are the creation and annihilation operators in real space $\hat{a}({\bf x})$ and $\hat{a}^{\dagger}({\bf x})$ satisfying
\begin{equation}
[ \hat{a}({\bf x}), \hat{a}^{\dagger}({\bf x}') ] = \delta ({\bf x} - {\bf x}')
\end{equation}
They are related to the standard operators in momentum space by
\begin{equation}
\hat{a}( {\bf x}) = \int \frac{dk}{(2 \omega_{\bf k})^{1/2}} e^{-i{\bf k}{\bf x}} \hat{a}({\bf k})
\end{equation}  
with $dk = \frac{d^3k}{(2 \pi)^3}$.
The normal- ordered Hamiltonian then reads
\begin{equation}
\hat{H} = \frac{1}{2} \int dx dx' \hat{a}^{\dagger} ({\bf x}) h({\bf x},{\bf x}') \hat{a}({\bf x}')
\end{equation}
 with 
\begin{equation}
h({\bf x},{\bf x}') = \int dk e^{-i{\bf k}({\bf x}-{\bf x}')} \omega_{\bf k}
\end{equation} 
while the Heisenberg - picture operator
\begin{equation}
\hat{a}({\bf x},t) = e^{i \hat{H}t} \hat{a}({\bf x}) e^{-i \hat{H}t} = 
\int dx' U({\bf x} - {\bf x}' ; t) \hat{a}({\bf x}')
\end{equation}
in terms of 
\begin{equation}
U( {\bf x}- {\bf x}';t) = \int dk e^{-i   {\bf k}( {\bf x}- {\bf x}')} e^{-i \omega_k t}
\end{equation}

\subsection{The coarse - graining operator}
Let us assume a cubic lattice with side length $L$ on a spatial hypersurface 
$\Sigma$ 
of Minkowski spacetime. The centers of the cubes are identified with  the points, coordinated by $( nL, mL, rL)$ with $n,m,r \in {\bf Z}$. Let us denote, by  $[.]: \Sigma \rightarrow \Sigma $ the function that takes a point ${\bf x}$ and assigns it at the point $[{\bf x}]$ of the center of the cube of the lattice in which ${\bf x}$ belongs. Also let $I({\bf x})$ denote the cube of the lattice in which ${\bf x}$ belongs.
\par
Our aim is to find a coarse graining operator, that intuitively does the following: to each field in spacetime it assigns a number of operators, which correspond to the smearing of the field over a  lattice cube and assumed to lie on its centre . It is rather easy to construct this kind of operator by considering the smeared fields.
\begin{equation}
{\bf P} \hat{a}(f) = \hat{a}(f^P)
\end{equation}
where $f^P$ is defined as follows
\begin{equation}
f^P({\bf x}) = \frac{1}{L^3} \int f({\bf x}') \chi_{I({\bf x})}({\bf x}') dx'
\end{equation}
where $\chi_S({\bf x})$ is the characteristic function of the set $S$.
It is easy to check that this definition produces an indempotent, self adjoint with respect to the Hilbert - Schmidt inner product operator on the space of observables \footnote{ This projector has a natural geometric significance which is worth stating. Recall that a field is an operator - valued distribution on $C_0(\Sigma)$, where $C_0(\Sigma)$ is the space of complex - valued functions of compact support on $\Sigma$. Now, the map $[.]$ is a projection operator, projecting $\Sigma$ to the lattice $\Sigma / L^3$. It is easy then to confirm that ${\bf P}$ is obtained from $[.]$ by two successive applications of the pull - back operation.}.
\par
In a formal sense one can write the operator acting on unsmeared fields 
\begin{equation}
{\bf P} \hat{a}({\bf x}) = \frac{1}{L^3} \int dx' P({\bf x},{\bf x}') \hat{a}({\bf x}')  
\end{equation}
where 
\begin{equation}
P({\bf x},{\bf x}') = \chi_{I({\bf x})}({\bf x}')
\end{equation}
If one then employs equation (2.7) we obtain the interesting result
\begin{equation}
{\bf P} \hat{a}({\bf x}) = \int \frac{dk}{(2 \omega_{\bf k})^{1/2}} e^{-i{\bf k}{\bf  [{\bf x}]}} \hat{a}({\bf k}) \rho_L({\bf k})
\end{equation}  
in terms of the function
\begin{equation}
\rho_L({\bf k}) = \prod_{i=1}^{3} \frac{\sin (k_i L /2)}{k_i L/2}
\end{equation}
This probably makes more transparent the meaning of the particular coarse - graining. The coarse grained field has a modifiacation in its spectral density and is essentially evaluated in the centers of the cubes in the lattice. It also implies the  useful expression
\begin{equation}
P({\bf x},{\bf x}') = \int dk e^{-i{\bf k}([{\bf x}]-{\bf x}')} \rho_L({\bf k})
\end{equation}

\section{The evolution equations}

Equipped with the above expression for the coarse graining operator, our task is now to write down an explicit form of the equation (2.3). This involves the evaluation of the noise term and of the kernel 
\begin{equation}
{\bf G}(\tau) = {\bf PL}({\bf 1} - {\bf P}) e^{i  {\bf L} \tau} ({\bf 1} -
{\bf P}) {\bf LP}
\end{equation}
We find it convenient to use the index notation employed in \cite{Ana}, where 
an ${\bf x}$ dependence of a kernel is denoted by an index and integration by contraction of the indices (for more details see Appendix A of this reference).
\par
It is easy to verify that
\begin{equation}
{\bf G}(\tau) \hat{a}^a = H^a{}_b U(\tau)^b {}_c \bar{H}^c{}_d \hat{a}^d
\end{equation}
where 
\begin{equation}
H^a{}_b = P^a{}_c h^c {}_d Q^d{}_b
\end{equation}
Also ${\bf Q} = {\bf 1} - {\bf P}$, $h$ is defined by equation (2.9) and a bar denotes complex conjugation.
Actually
\begin{eqnarray}
H({\bf x},{\bf x}') = \int dk dk' \rho({\bf k})  e^{-i{\bf k}[{\bf x}]+i{\bf k}'{\bf x}'} \omega_k 
\nonumber \\
\times \left( (2 \pi)^3 \delta ({\bf k}-{\bf k}') - \rho_L ({\bf k}') \Delta ({\bf k}',{\bf k})\right)
\end{eqnarray}
where 
\begin{eqnarray}
\Delta({\bf k},{\bf k}') = \int dy e^{-i {\bf k} [{\bf y}] * i {\bf k}'{\bf y}} \nonumber \\
= \rho_L({\bf k}') \prod_{i=1}^3 \frac{ 1 - [(k_i - k'_i)L]^2}{1 + [(k_i- k'_i)L]^2 }
\end{eqnarray}
Substituting this together with (2.11) in the above equation we obtain for the kernel ${\bf G}$
\begin{eqnarray}
G({\bf x},{\bf x}';\tau) = \int dk dk' dk'' e^{-i {\bf k} [{\bf x}] + i {\bf k}'' [{\bf x}']} \omega_k \omega_{k'}
\rho_L({\bf k}) \rho_L({\bf k}') e^{-i \omega_{{\bf k}'} \tau}
 \nonumber \\
 \times    \left( (2 \pi)^3 \delta({\bf k} - {\bf k}')
- \rho_L({\bf k}') \Delta({\bf k}',{\bf k}) \right) \left( (2 \pi)^3 \delta({\bf k}'' - {\bf k}')
- \rho_L({\bf k}') \Delta({\bf k}',{\bf k}'') \right)
\end{eqnarray}

The same steps would hold for the calculation of the noise terms. The corresponding kernel is easily found to be
\begin{eqnarray}
F({\bf x},{\bf x}';t)  = \int dk dk' dk'' e^{-i {\bf k} [{\bf x}] + i {\bf k}'' {\bf x}'} \omega_k 
\rho_L({\bf k}) e^{-i \omega_{k'} t}
 \nonumber \\
 \times    \left( (2 \pi)^3 \delta({\bf k} - {\bf k}')
- \rho_L({\bf k}') \Delta({\bf k}',{\bf k}) \right) \left( (2 \pi)^3 \delta({\bf k}' - {\bf k}'')
- \rho_L({\bf k}'') \Delta({\bf k}'',{\bf k}') \right)
\end{eqnarray}
Now this equation together with its conjugate for $\hat{a}^{\dagger}$ are sufficient to determine the hydrodynamic behaviour of the expectation values of the smeared fields
fields. (Recall that free evolution does not mix the n-point function levels). Denote the coarse- grained Hamiltonian  kernel by $\tilde{h}({\bf x}-{\bf x}')$ .
\begin{equation}
\tilde{h}({\bf x},{\bf x}') = \int dk dk'\rho({\bf k})  e^{-i{\bf k}[{\bf x}]+i{\bf k}'{\bf x}'} \omega_k 
 \rho_L ({\bf k}') \Delta ({\bf k}',{\bf k})
\end{equation}
In terms of the above kernels the relevant evolution equations are:
\begin{equation}
i \frac{\partial}{\partial t} a({\bf x},t) - \int dx' \tilde{h}({\bf x},{\bf x}') a({\bf x}',t)
+ i \int_0^t d \tau G({\bf x},{\bf x}';\tau) a({\bf x}', t - \tau) = \xi ({\bf x}, t)
\end{equation}
and its complex conjugate. ( We have denoted $a({\bf x}) = Tr \left( \hat{\rho}(0)
 {\bf P} \hat{a}({\bf x}) \right) $).
 The noise can be modelled by the knowledge of 
its correlation functions. For example:
\begin{eqnarray}
\langle \xi({\bf x}, t) \rangle = \int dx F({\bf x},{\bf y}) \langle \hat{a}({\bf y}) \rangle_0 \\
\langle \xi({\bf x}, t) \xi({\bf x}', t') \rangle = \int dy dy' F({\bf x},{\bf y};t) F({\bf x}', {\bf y}';t') 
\langle \hat{a}({\bf y}) \hat{a}({\bf y}') \rangle_0 \\
\langle \xi({\bf x}, t) \bar{\xi}({\bf x}', t') \rangle = \int dy dy' F({\bf x},{\bf y};t) \bar{F}({\bf x}', {\bf y}';t') 
\langle \hat{a}({\bf y}) \hat{a}^{\dagger}({\bf y}') \rangle_0 \\
\end{eqnarray}
and continuing for the higher order correlation functions.

\section{ The Gaussian approximation}
So far our approximations have been purely formal in the sense that no physical input has entered our equations. They have all been derived by follwing the mathematics implicit in the Zwanzig technique. We still need to enforce the three conditions we have stated in the introduction, in order to get meaningful hydrodynamic equations.
\par
The first observation is that the expectation value of $\xi({\bf x},t)$ is dependent solely on the expectation value of the field at the initial moment of time. 
Now, the scale by which $\langle \hat{a}({\bf x}) \rangle$ varies is exactly what should be identified with the length $l_{mic}$. Assumption (2) implies that with good approximation we can set is to zero. Hence 
\begin{equation}
\langle \xi(t,{\bf x}) \rangle \simeq 0
\end{equation}
Note that this condition does not imply, that the initial state has no irrelevant component. What is rather implied, is that the part of the state, where the one - point functions have support lies in the eigenspace of ${\bf P}$. In particular higher order correlation lengths in the initial state can be larger than $L$ and these are the ones that give rise to the noise \footnote{ Since a quantum field is a system with an infinite number of degrees of freedom, in any state there is {\it a priori} an infinite number of length parameters, subsets of them are roughly governing the behaviour of $n$-point functions. Even a Gaussian state is characterized by at least two such scales: the typical length scale of its mean value and the correlation length read from its two - point function, which are generically quite distinct.}. Note, that this assumption for the initial condition essentially fixes the origin of the time axis.
\par
Now, the difficulty of obtaining simple hydrodynamic equations is mathematically connected to two facts: \\
1. The complicated expression for the function $\rho_L$. \\
2. The appearance of the  $[.]$ in our exponential, which also gives rise to the presence of the function $\Delta$.\\
What we would like to have is to substitute $[{\bf x}]$ with ${\bf x}$ in the exponential and $\rho_L$ with another easier to handle function. The condition for the former  is essentially the possibility of taking the continuum limit for the lattice, which essentially amounts to the condition $ L << l_{obs}$. And we shall show that the later substitution possible by the use of the other hydrodynamic assumption $l_{mic} << L$.
\par
The idea is instead of using the characteristic function for the set $I({\bf x})$ in equation (2.14)
 to use a smeared characteristic function. The most efficient way to perform this is to substitute the kernel $P({\bf x},{\bf x}')$ with the Gaussian
\begin{equation}
P({\bf x},{\bf x}') = \frac{1}{(\pi)^{3/2} L^3} e^{-({\bf x} - {\bf x}')^2/L^2}
\end{equation}
Note that this substitution involves essentially two steps: \\
1. Using a smeared characteristic function, which can be justified only if the width is much larger than the scale of variation for the  expectation value 
of the field field, hence $l_{mic} << L$. The error will be then of the order
$(l_{mic}/L)^2$.
\\
2. Substituting $[{\bf x}]$ with ${\bf x}$, which is equivalent as we explained to the invocation of a continuum assumption. \\
At this point we should stress that $P({\bf x},{\bf x}')$ does not correspond to a true projector, since it is not indempotent. But in the particular subspace of the field states, that the hydrodynamic conditions of separations of the lengths is satisfied, it is very close to  a true projector, with an error of the order 
$\max \{(l_{mic}/L)^2, L/l_{obs} \}$.
\par
Hence equation (2.17) can now be writen in the form
\begin{equation}
{\bf P} \hat{a}({\bf x}) = \int \frac{dk}{(2 \omega_k)^{1/2}} e^{-i{\bf k}{\bf x}} \hat{a}({\bf k}) \tilde{\rho}_L ({\bf k})
\end{equation}
where 
\begin{equation}
\tilde{\rho}_L({\bf k}) = e^{- L^2 {\bf k}^2/4}
\end{equation}
and the function $\Delta$ becomes approximately
\begin{equation}
\Delta ({\bf k},{\bf k}') = (2 \pi)^3 \delta( {\bf k} - {\bf k}')
\end{equation}
Hence the kernels governing the evolution of the smeared fields simplify significantly
\begin{eqnarray}
G({\bf x},{\bf x}' ;\tau) = \int dk e^{-i{\bf k}( {\bf x} - {\bf x}')} \omega_k^2 [ ( \tilde{\rho} - \tilde{\rho}^2)({\bf k})]^2
e^{-i \omega_k \tau}
\\
F({\bf x},{\bf x}' ; t) = \int dk e^{-i{\bf k}({\bf x} - {\bf x}')} \omega_k e^{-i \omega_k t} \tilde{\rho}({\bf k}) \left( 1 - \tilde{\rho}({\bf k}) \right)^2
\\
\tilde{h}({\bf x},{\bf x}') = \int dk e^{-i{\bf k} ({\bf x} - {\bf x}')} \tilde{\rho}^2 ({\bf k}) \omega_k
\end{eqnarray}
Now the kernels $G$ and $\tilde{h}$ act on the coarse grained expectation values of the field, in which the high energy modes are suppresse by the Gaussian smearin. Hence within an approximation of $(L/l_{obs})^2$ we can keep only the lowest order in the expansion of $\tilde{\rho}_L$ with respect to $L$. This gives
\begin{eqnarray}
\tilde{h}({\bf x},{\bf x}') = h({\bf x},{\bf x}') \\
G({\bf x}, {\bf x}'; \tau) = \frac{L^4}{16} \int dk {\bf k}^4 \omega_k^2 e^{-i{\bf k}({\bf x} - {\bf x}')} e^{-i \omega_k \tau} \nonumber \\
= \frac{L^4}{16} \nabla^4 (m^2 + \nabla^2) U({\bf x}, {\bf x}', \tau)
\end{eqnarray}
On the other hand $F$ acts on the unsmeared fields, hence we cannot drop the exponential damping term. But using the same argumentation we can expand $(1 - \tilde{\rho}_L)$
to obtain
\begin{eqnarray}
F({\bf x}, {\bf x}' ; t) = \int dk e^{- i {\bf k} ({\bf x} - {\bf x}')} \omega_k e^{- L^2 {\bf k}^2/4} e^{- L^2 {\bf k}^2/4} \nonumber \\
= -i \frac{L^2}{4} \frac{\partial}{\partial t} \nabla^2 \tilde{U}({\bf x},{\bf x}')
\end{eqnarray}
with 
\begin{equation}
\tilde{U}({\bf x},{\bf x}') = \int dk e^{- i {\bf k} ({\bf x} - {\bf x}')} \omega_k e^{- L^2 {\bf k}^2/4}
\end{equation}
which is essentially the relevant part of the evolution operator.
Hence the kernels giving rise  to dissipation and noise can be writen in terms of the full and the relevant part of the evolution operator.

\section{Remarks}
Our main results is essentially the Langevin type of equations (3.9) - (3.12) written in terms of the kernels (4.10) and (4.11). They should form the basis of construction of hydrodynamic equations, since having assumed classicality one can use the classical expressions to write energy and momentum density in terms of $a({\bf x})$ and $a^*({\bf x})$.  
\par
The equations admit simplifications in various regimes. For instance the dissipation term for a massless field reads simply
\begin{equation}
i \nabla^6   \int dx' \dot{a}({\bf x}', t - |{\bf x} - {\bf x}'|)
\end{equation}
where the non- local in time behaviour is encoded in the redarded time dependence of $a$. But, it has been impossible for this choice of coarse - graining to obtain a local in time equation, which could lead   a field counterpart of the Navier - Stokes equation.
\par
One reason for this is probably to be found in our choice of coarse graining. While physically realistic, it is conceivable that it might not be the most natural  choice. The formalism of generalised quantum mechanics allows in principle space-time coarse grainings and it is probably through them that a maximal 
quasiclassical domain can be constructed. The possibility that a temporal coarse - graining is needed as well, is reinforced by examining standard non- relativistic applications of the projector technique. The timescale of observation ought to be taken much larger than the characteristic time parameter appearing in the kernel ${\bf G}$, if we want to obtain local in time (or Markovian) evolution equations \cite{ReMi}.
\par
As far as the noise is concerned, it is highly unlikely that it can be approximated by white noise, even though for particular initial state the Gausian approximation might be a reasonable assumption. One needs to make physical assumptions for the initial state, particular to the problem in hand \footnote{ Note that in the projector technique an ignorance interpretation is assumed for the initial density matrix, that is the relevant component only can be known and the irrelevant should be modelled, usually by conjunction of physical arguments and 
application of the minimum entropy principle. While this is not a problem for most physical systems, in cosmology (where there is no  meaning of an ensemble of systems) one has to be extremely careful on the assumptions involved for the modeling of noise}.
\par
There are a re three interesting points to make, concerning our approach \\
1. For our particular class of initial states, the contribution of noise and dissipation is very weak (the coupling constant is essentially $L^4$). Hence, while for short times one can safely rely on the evolution governed by the self dynamics of the coarse - grained field, as time increases  predictability is going to be degraded, the faster the larger the long- range correlations in the initial state are. One would eventually have to use statistics (classical), constructing equations for the probability densities of the smeared field.
\\
2. The parameter governing both the loss of predictability and the flow of energy due to dissipation is the phenomenological coarse -graining  parameter $L$. It is one of the great appeals of the histories approach, that they give an algorithm (computing maximal decohering histories), following which it can be determined. In our considerations we have encountered a feature, that could potentially augment our understanding of the quantum to classical transition: {\it The parameters determining minimum coarse - graining such that decoherence is possible, will appear among the coupling constants determining the strength of 
dissipation and diffusion in any hydrodynamic description}
\\
3. In our coarse graining, essentially the low energy modes contribute to the relevant observables, but high energy modes (even if not directly coupled to the low energy modes) induce a significant noise in the equation of motion, by virtue of possible long - range correlations of the initial state.
\par
We should finally remark, that a more promising approach towards investigating spacetime coarse grainings, implemented directly at the level of the field 's generating functional is currently under investigation.

\section{Aknowledgements}
I would like to thank J.J. Halliwell and E: Calzetta for discussions on relevant issues.  The research was supported by Generalitat de Catalunya by grant 96SGR-48.

\end{document}